\documentclass[10pt,conference]{IEEEtran}

\usepackage[english]{babel}
\usepackage[utf8]{inputenc} 
\usepackage[T1]{fontenc} 

\usepackage[stretch=10]{microtype} 

\usepackage{cite}
\usepackage{amsmath,amssymb,amsfonts}
\usepackage{mathtools}
\usepackage{algorithmic}
\usepackage{graphicx}
\usepackage{textcomp}

\usepackage{subcaption}
\usepackage{float}

\usepackage[table, svgnames]{xcolor} 
\usepackage{booktabs}
\usepackage[hyphens]{url}
\usepackage[breaklinks=true, colorlinks=true, allcolors=blue, pdfencoding=auto, psdextra, pagebackref, linktoc=all]{hyperref}
\usepackage{bookmark}
\ifdefined\backref
\renewcommand*{\backref}[1]{}
\renewcommand*{\backrefalt}[4]{%
	\ifcase #1 (Not cited.)
	\or  (Cited on page:~#2)
	\else  (Cited on pages:~#2)
	\fi%
}
\else
\fi

\usepackage{siunitx} 
\usepackage[autostyle=true,english=american]{csquotes} 
\MakeOuterQuote{"}

\def\BibTeX{{\rm B\kern-.05em{\sc i\kern-.025em b}\kern-.08em
 T\kern-.1667em\lower.7ex\hbox{E}\kern-.125emX}}
\usepackage{algorithm}
\usepackage{algorithmic}
\usepackage{todonotes} 
\usepackage{cleveref} 
\crefname{lstlisting}{listing}{listings} 
\Crefname{lstlisting}{Listing}{Listings} 
\Crefname{figure}{Fig.}{Fig.}
\Crefname{equation}{Eq.}{Eq.}

\usepackage{soul}
\sethlcolor{WhiteSmoke}

\DeclarePairedDelimiter\cardinality{\lvert}{\rvert} 

\begin{document}

\title{Automatic Error Classification and Root Cause Determination while Replaying Recorded Workload Data at SAP HANA

}

\author{\IEEEauthorblockN{Neetha Jambigi}
\IEEEauthorblockA{University of Innsbruck \\
Innsbruck, Austria }
\IEEEauthorblockA{neetha.jambigi@student.uibk.ac.at}
\and
\IEEEauthorblockN{Thomas Bach}
\IEEEauthorblockA{SAP \\ Walldorf, Germany\\
0000-0002-9993-2814}
\and
\IEEEauthorblockN{Felix Schabernack}
\IEEEauthorblockA{ 
SAP\\ Walldorf, Germany\\felix.schabernack@sap.com
}
\and
\IEEEauthorblockN{Michael Felderer}
\IEEEauthorblockA{University of Innsbruck \\
Innsbruck, Austria \\
0000-0003-3818-4442}
}

\maketitle

\begin{abstract}

Capturing customer workloads of database systems to replay these workloads during internal testing can be beneficial for software quality assurance. However, we experienced that such replays can produce a large amount of false positive alerts that make the results unreliable or time consuming to analyze. Therefore, we design a machine learning based approach that attributes root causes to the alerts. This provides several benefits for quality assurance and allows for example to classify whether an alert is true positive or false positive.
Our approach considerably reduces manual effort and improves the overall quality assurance for the database system SAP HANA. We discuss the problem, the design and result of our approach, and we present practical limitations that may require further research.


\end{abstract}


\section{Introduction}
\label{s:introduction}

As SAP develops software to manage business operations, the quality of the software is an important aspect. Customers demand that the software works as expected. Even more, in the context of business operations, customers expect that newer versions of a software provide at least the same functionality as the version before, i.e. they expect that new versions do not contain regressions. This requirement is rather challenging for very large software projects as each modification to the source code can potentially lead to unexpected side effects that may not be predictable by humans due to the project size.

In the case of SAP HANA, a database management system for enterprise applications developed by SAP~\cite{FarberHANA2012journal, FarberHANA_SIGMOD2012}, the source code consists of about 10 million lines of code and is changed about 200 times a day, making it impossible for a single person to understand all parts and all modifications at any point of time. Therefore, SAP maintains an extensive set of over \num{100000} regression tests, i.e. tests that assure a new software version has no regressions~\cite{bach:2017:coverageBasedReduction, Bach:2018:EconomicApproachTestStages}. However, even with such an extensive test suite, customers may use the software in a way that is not explicitly tested. For example, customers may use different table and data distribution, workload management, system utilization, system size or system configuration. It is therefore important to test a new software version with customer workloads. One approach at SAP to achieve this is recording all workload from a customer system over a certain time~\cite{baek2020rsx}. This workload can then be replayed at any time for any new version. On a high level, the implicit test oracle, i.e. the verification whether the test has passed or not, is whether the replay executes successfully or if there are any errors encountered during the execution. However, the recording also allows a fine-grained analysis of individual queries and performance. Capturing database workload for testing purposes is also investigated by previous work~\cite{snowflake2018dbtest, arruda2016capture, oracleDatabaseReplay2008, wang2009real}. Other related fields are capture and replay approaches for GUI~\cite{sprenkle2005automated} and selective replays for program executions~\cite{orso2005selective}, but having the direct connection to the database provides more information compared to a scripted testing type that is often applied for GUI testing. In addition, capturing customer workloads has also benefits compared to artificial tests.

While the idea of replaying recorded database queries sounds simple in theory, it poses severe challenges in practice. Besides the size of such replays, the costs to record and re-run them or legal issues with respect to customer data that must be clarified, we also found that this approach shows false positives, i.e. errors that in fact do not indicate regressions. Such false positives are caused by non-deterministic behavior such as multi-threading and concurrency, sources of randomness, date and time, missing functionality of the record tool, hardware specific code paths, influence of the record tool on the observed behavior, missing data, different configuration settings. Another category of sources for false positives is external factors such as infrastructure issues with the network, hardware, or the storage application. Altogether, we found in practice a wide range of reasons why false positives appear.

Given that false positives appear, we need a strategy to incorporate them in the test assessment. Assuming a replay shows a set of $P_a$ errors, then we can classify all shown errors in two distinct subsets $P_a = P_t \cup P_f$ where $P_t$ are true positive errors representing regressions and $P_f$ are false positive errors, i.e. no regressions. It is desirable to keep $P_f$ empty, but it is in practice not possible to achieve this with a reasonable effort. Analyzing each item in $P_f$ by developers can easily create huge efforts if $\cardinality*{P_f} \ge 100$ and may not be feasible if $\cardinality*{P_f} \ge 10^5$. In practice, we observed  $\cardinality*{P_f} \ge 10^6$ for large and complex replays due to inherent characteristics and functional limitations of the capture and replay.

Handling a large set of $P_f$ therefore requires an automated approach. As the items in $P_f$ may have a common structure, a machine learning approach seems to be well suited to classify the issues automatically. For this purpose, we identify a set of information about each error such as the related SQL queries, error codes, error messages, stack traces of errors or information from bug tracker and project management tools. In the machine learning domain, we call such characterizing information \emph{features}. Given a set of pre-defined error classifications, the machine learning approach derives a mathematical model of how to conclude the error classification based on the features, i.e. the algorithm \emph{learns} to classify.

Furthermore, our approach does not only do a binary classification between true positive and false positive, it also learns to associate the root causes for each item. With that, we can group the true and false positives results into common sets to further support the analysis of the test results. For this purpose, we also maintain a database of previously encountered and manually verified testing results.

Other works also investigate the classification of errors into true positives and false positives in the domain of software defect~\cite{podgurski2003automated, kahles2019automating, feng2018empirical}, the analysis of logs~\cite{du2017deeplog, bertero2017experience} or, more general, in outlier analysis~\cite{hodge2004outlierSurvey}. In contrast, we utilize additional information we gain from the database itself to improve the accuracy of our classification. Other work also utilizes the information of defect tracking tools and machine learning for automatically prioritizing bugs~\cite{uddin2017survey}, detecting bug duplication~\cite{zou2016duplication, neysiani2020automatic} and classification of root causes for bugs~\cite{catolino2019not}. In our case, we combine bug categorization with classification supported by the additional information we gain from the database during the execution and from previous classification that were verified by human operators.




This paper is structured as follows. \Cref{s:tool_design} presents the tool design and the industrial context of our approach. In \Cref{s:MLComponent} we describe our machine learning approach to error classification and present a detailed account of different stages and results of evaluations in \Cref{s:Evaluation}. In \Cref{s:ApplicationDiscussion} we discuss the application in practice and feedback from the end users. \Cref{s:challenges} contains our experiences, limitations, lessons learned and open questions. We conclude in \Cref{s:conclusion}.

\section{Industrial Context and Tool Design}
\label{s:tool_design}


As described in \cref{s:introduction}, the development of SAP HANA utilizes a large set of unit and integration tests, but testing with customer workloads is still a beneficial task for quality assurance. Replaying captured customer workload can help to identify software defects early that are not detected by other tests types. In comparison to other scripted tests, there is no artificial data or underlying assumptions made by test engineers that can prevent the software system from malfunctions. Additionally, we can selectively record and replay critical periods in the database system driven by specific application processes and user behavior such as year-end closing, financial planning, or weekly reports. However, compared to artificial tests, the replay tests may be less stable. To ensure that test and quality owners can trust the overall product quality assessment provided by the test results before releasing a new database version to customers, it is essential to distinguish between the true and false positives automatically.

During the replay of a captured workload, we collect the result of the execution and a set of attributes for each query (typically a SQL statement). Such a collection represents an event. Operators are the quality assurance experts responsible for the replay result analysis. Operators assess the collected events to identify the root causes of the failures. In this section, we briefly describe how a rule-based approach supports the assessment of such events and why such an approach has severe limitations. We then introduce our machine learning based tool to automatically support the assessment, which we developed under the internal name MIRA - Machine Intelligence for Replay Analysis. 

\subsection{Limitations of the Rule-Based Approach}
 Currently, at SAP, a solution for automating the replay error reasoning involves utilizing a combination of Request Id and Error Code. Request Id and Error Code are a part of the set of attributes collected for an event. There are Request Ids to identify failed replay events. There is an associated Error Code for every failed event in the replay. When this combination for a specific capture appears for the first time, it gets tracked as a failure. An operator assigns a tag value to the failures depending on the root cause, that indicates whether the failure is a real issue or not. If applicable, a bug id is assigned as well. In a subsequent replay, if the tool detects a known tuple for the (Request Id, Error Code) combination, the failed event is labeled based on the tag of a matched entry. The tag assignment for failed events is done manually once for every capture to create a baseline to compare in successive replays.
 
This approach has the following limitations:
i) Different captures could be workloads captured from the same system at different periods or may belong to different systems altogether. Request Ids are automatically generated for every SQL statement when a workload is captured and is a unique identifier for an SQL query. It is used to identify a particular query in replays of the same capture. For example, Query Q: \textit{SELECT col1, col2 FROM Table\_A;} has a Request Id 101 in capture 1. In every replay of the workload from capture 1, the Query Q will have the same Request Id 101. However, if the same query is executed in different captures, it will be assigned a different Request Id which is valid only within the replays of the respective capture. When Query Q fails across two different captures it is tracked as Event A and Event B as shown in \Cref{tab:Rulebased}. Even though failed events A and B have the same error codes, the tag of event A cannot be used to assign a tag for event B due to different Request Ids. Therefore, even though similar errors occur, we cannot compare failures across different captures.
ii) There can be discrepancies in label assignment between different operators for errors arising due to similar reasons. 
iii) In HANA, there are cases where the same error code encompasses multiple types of errors. This problem becomes clear when the relevant error messages are analyzed. Therefore, comparing merely error codes is not adequate.


\begin{table}[ht]
\centering
\caption{Rule-based approach}
\label{tab:Rulebased}
\begin{tabular}{ccccr}
\toprule
Capture & Event Id & (Request Id, Error Code) & Tag & Bug Id \\ \midrule
1  & A  & (101, 202)     & Bug & 1234 \\
2  & B  & (102, 202)     & Bug & 1234 \\
\bottomrule
\end{tabular}
\end{table}



\subsection{The MIRA Tool}



\begin{figure*}[ht]
 \centering
 \caption{Overview of MIRA within the Capture and Replay test pipeline of SAP HANA}
 \includegraphics[scale=0.80]{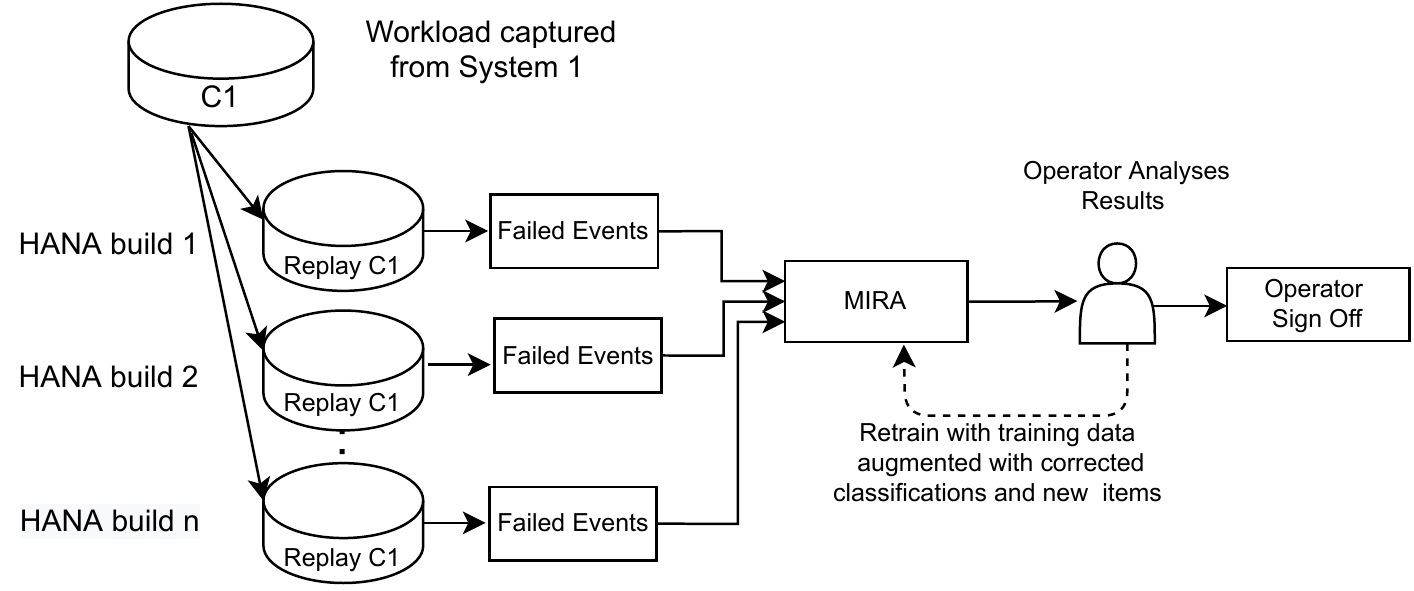}
  \label{fig:miraBlockDiagram}
\end{figure*}

 MIRA is developed and integrated as a microservice into the pipeline associated with automated capture and replay as a part of the release and test process. Whenever a captured workload is replayed, results are automatically forwarded to MIRA for the classification of failed replay events. The data, we train MIRA with, will change due to new software defects being constantly introduced in the development process or due to changes in the capture and replay tool itself. The machine learning model will need to adapt to the changes of the data. The architecture and interface of the microservice are built to easily accommodate such changes. The service offers a machine-learning-process-centric API that allows control of an end-to-end machine learning process from creating a prediction to correcting predicted classifications and re-training these corrections back into the model. In \Cref{fig:miraBlockDiagram} we present an overview of MIRA's position in the  capture and replay test process, where a capture C1 is replayed in multiple HANA versions and the failed events are redirected to MIRA for classification. We have implemented this API in close cooperation with the operators to incorporate their feedback and improve MIRA's usability as a framework. 
 

\section{Machine learning based Error Classification for Replay Analysis in MIRA }
\label{s:MLComponent}
As described in the previous section, the core part of MIRA is a machine learning based solution to classify the failed replay events into the root causes of their failures. There are two main aspects we address as a part of our solution: i) Classification Model: Learning a model to classify failed events into their root causes. ii) Uncertainty Measures: Defining measures to reflect the uncertainties of the model.

\subsection{Classification Model}In MIRA, we only use failed events to train a classification model. Amongst all the collected attributes for the failed replay events, we use a subset that includes HANA Server Error Code, HANA Server Error Message, Request type, SQL Type and SQL Sub Type to train a classifier. Based on our domain knowledge, these attributes are the most predictive of the reason for the failure. In our data, all the selected attributes are categorical, apart from the error message. All categorical and unstructured data need to be converted into numerical data for utilizing them as an input to the machine learning algorithms. There are very few ways to encode categorical attributes with a categorical target. Furthermore, each of the attributes contains several unique values in them and one hot encoding will result in very high dimensional data. Discretization or grouping the values of an attribute can help reduce the number of unique values and thus reduce the number of resulting dimensions. However, this is not possible in our data as the attributes have no discernible ordinality within them. 

The error messages in our data are unstructured and have different lengths. We treat every error message as a document and employ techniques like  TFIDF~\cite{salton1986introduction} and Doc2Vec~\cite{le2014distributed} to create a vectorized representation of the error message of an event. Concatenating the one hot encoded event attributes with the vectorized error message further exacerbates the high dimensionality problems.
 
 We apply the K-Nearest Neighbors (KNN) algorithm as the classification model in MIRA. KNN relies on a distance metric to identify neighbors or the most similar instances of an instance to be classified. The curse of dimensionality is a well-known problem in machine learning and is known to adversely affect classifiers like the KNN~\cite{mucherino2009k, pestov2013k}. To address the problem of encoding our data while avoiding high dimensions, we introduce a distance function detailed in \Cref{alg:customdistance}.  We calculate a total distance between two events by adding up the distances between the individual attributes of the events. We refer to this distance as 'custom distance' (CD) in the paper. With custom distance, we can assign different weights to the attributes. The weight of an attribute is multiplied with the distance of the attributes resulting in increased or decreased influence of the attribute towards the distance calculation. For instance, in the production model of MIRA, Error Codes and Error Messages have higher weights in comparison the other attributes as they are the most effective at identifying the reason for the failure. We set the weights of the attributes in the production model based on cross-validation results and domain knowledge of our experts.

\begin{algorithm}
\caption{CustomDistance}
\begin{algorithmic}[1]
\STATE \textbf{Input:} $x, y \leftarrow Event$
\STATE $distance \leftarrow 0$
\FOR {$i \gets 1$ to $len(x)$}
\IF{$type(x[i]) ==\ 'ErrorMessage'$}
\STATE $x[i] \leftarrow TextVectorizer(x[i])$
\STATE $y[i] \leftarrow TextVectorizer(y[i])$
\STATE $distance \leftarrow distance + cosine\_distance(x[i], y[i])$
\ELSE
 \STATE $binary\_comparison \leftarrow 0$
 \IF {$x[i] != y[i]$}
 \STATE $binary\_comparison \leftarrow 1$
 \ENDIF
 \STATE $distance \leftarrow distance + binary\_comparison$
\ENDIF
\ENDFOR
\RETURN $distance/len(x)$
\end{algorithmic}
\label{alg:customdistance}
\end{algorithm}

\subsection{Uncertainty Measures}To make MIRA reliable, we identify mechanisms that reflect uncertainties of the model. For this purpose, we augment every classification in MIRA with two measures that reflect the classifier’s confidence in classifications. We refer to them as Probability and Confidence. Both the measures range between 0 to 1. We consider a classification unreliable if the calculated probability or confidence has a low value. The threshold for low values is a pre-defined constant based on our experiments with the training data. Operators manually inspect and correct such unreliable classifications. The corrected classifications are used to train MIRA further.

\textit{Probability}: We calculate the probability for KNN on voting weighted by the distance of the neighbors. This reflects uncertainty when there are members from different classes contributing equivalently towards the classification. However, the probability alone does not indicate uncertainty in cases where a new event without resemblance to the events in the training data is classified with a probability of 1.0. This occurs when the neighbors contributing towards a classification all belong to the same class but are almost entirely different from the event to be classified.

\textit{Confidence}:
Confidence is an additional measure to identify uncertainties not reflected by the probability. By design, custom distance allows us to calculate the maximum possible distance between two vectors based on weights assigned to the features.  

To calculate the confidence of a classification, we adopt a modified version of KNN confidence criteria as defined in previous work~\cite{dalitz2009reject} (Eq. (5)). The modification is with respect to the boundary consideration for the classification. In our case, we use the \emph{maximum distance} between two event vectors, calculated using custom distance described in \Cref{alg:customdistance}, as a boundary. 
 The confidence for a predicted class C for an event x is calculated as shown in \Cref{eqn:confidence}: 
\begin{equation}
 f(x, C) =\left ( 1 - \frac{min\{d(x, y_j )|\ 1<j<k \ and\ y_j \epsilon C \}}{maximum\ distance} \right )
\label{eqn:confidence}
\end{equation}
 where $d(x, y)$ is the custom distance between events $x$ and $y$. The numerator is the distance from event x to the nearest neighbor from class C belonging to the K neighbors contributing towards the classification. The denominator is the maximum possible distance between two event vectors with custom distance. The distance between two events is at its maximum when there are no matching values in any of the selected attributes. Therefore, we consider two events to be considerably different from one another when the distance between them exceeds a pre-defined threshold for the distance. 
 
 
 
 
\section{Evaluation}
\label{s:Evaluation}
In this section we present an overview of our data, data processing, evaluation procedure, and the results of our approach. We perform a $k$-fold stratified cross-validation~\cite{stone1974cross} for hyperparameter optimization and classifier evaluation. Since we attempt to handle the issue of high dimensionality with the Custom Distance (CD), we compare our approach to KNN with Euclidean Distance (ED) \cite{o2006elementary}, a distance measure that suffers in high dimensions. Additionally, we choose XGBoost\cite{Chen:2016:XST:2939672.2939785} as it has been shown to be a very effective classification algorithm across several domains of machine learning \cite{bentejac2021comparative}. We one hot encode all the categorical attributes, and concatenate them with vectorized error message to prepare the data for the baseline models. We present results for both TFIDF and Doc2Vec vectorization techniques. \Cref{tab:FinalHyperparameters} shows the final hyperparameter settings. We present both F1-scores weighted by class size (F1 Weighted) and a simple average across all classes (F1) in the results. We consider this representation suitable for data with a class imbalance as in our dataset.

\begin{figure*}[ht]
\centering
 \caption{Overview of evaluation process in MIRA }
 \includegraphics[scale=0.65]{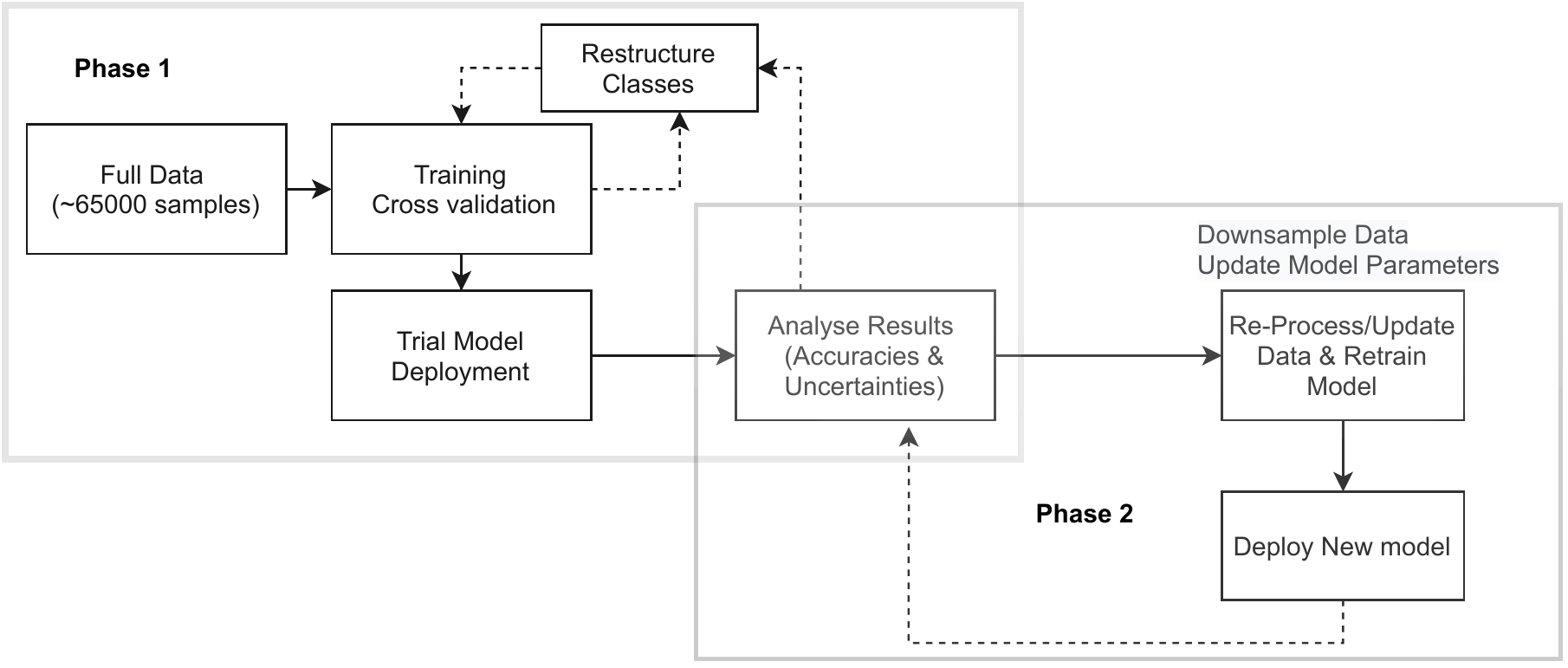}
  \label{EvalOverview}
\end{figure*}

\subsection{Data}
Prior to MIRA, operators investigated each failed event manually. If the failure was caused by a known issue, operators assign a reason for it as part of the replay analysis. Otherwise, a new bug is created, linked to the failure reason, and assigned to the event. The reason describes the fundamental cause of a failure, including reported HANA bugs, JIRA issues, and capture and replay problems. These failed events are used as MIRA training data. Discrepancies such as failed events with the same root cause but different labels occur as a result of different operators analyzing replays or mistakes made during replay analysis. Our domain experts reviewed it for labeling errors to ensure that the training data was of high quality.


\Cref{tab:event_table} presents some examples of the data where each row represents one failed event. We categorize failures based on their root cause. Some tables, for example, are unavailable during the replay due to failure to install specific HANA plugins during the replay. Attempts to access such tables are likely to fail with errors such as "Could not find table/view X". These failures are regarded as false positives because the plugins are not available for the specific HANA binary. They are grouped under a single root cause that represents the failure to install HANA plugins. As a result, several thousands of failures can be aggregated under only a few hundred root causes. Furthermore, the set of root causes grows as HANA evolves and new issues are identified. We currently have a total of 93 classes of which we categorize 73 as false positives and remaining 20 classes as real database issues. Each class represents one root cause or reason of failure. The attribute Type of SQL indicates whether the statement is a DDL, DML, etc., SQL Sub Type indicates finer granularity of the SQL type. The Error Code and Error Message are collected for every failed event from the HANA SQL interface exposed to HANA clients interacting with the database. There are 63 unique Error Codes, 5 SQL types, 16 SQL sub types, and 9 types of Request Names in the data.
\begin{table*}[ht]
\centering
\caption{Data example with artificial content for failed events and labels}
\label{tab:event_table}
\begin{tabular}{rrlrrl|l}
\toprule
Event Id & Error Code & Error Message     & SQL Type & Subtype & Request Type & Reason \\ \midrule
10000021 & 250986   & 'invalid argument: cannot determine volume id'   & 1    & 1    & Type1  & Reason1  \\
10000022 & 7000012  & 'Internal error: error=invalid table name: Could not find table' & 7    & 1    & Type2  & Reason2  \\ 
\bottomrule
\end{tabular}
\end{table*}

\label{classimbalance}
\subsubsection{Class Imbalance} Considering that some failures occur more frequently than others, there are more samples for some types of failures compared to others creating a class imbalance in the data. We handle the class imbalance in the dataset to some extent by downsampling the larger classes. We cannot upsample the minority classes through duplication as it could allow unimportant terms like temporary names to influence the vectorization of Error Message. Creating artificial samples from our data is quite challenging, as it contains just categorical and textual attributes.
 
 Random downsampling is not feasible for most of the classes in our data because a single root cause might consist of multiple errors that vary slightly in pattern and it is necessary to retain a representation of all the patterns in the data. To this end, we refine the dataset one class at a time. We collected all known error patterns within each class and selected enough samples from each pattern to have an effective representation of the error within the data. We performed cross-validation after each class is processed, to assess impacts of downsampling on the whole classification results. We started with a dataset containing about \num{65000} training samples, which was reduced to about \num{25600} samples after the downsampling process. \Cref{fig:ClassDist} shows the final class distribution. The downsampling also eliminates a substantial amount of unimportant terms from the vocabulary of the resulting Doc2Vec and TFIDF vectorization models. The vector size in TFIDF reduced to about \num{6000} from about \num{14000}  and the number of unique terms in the Doc2Vec model also reduced. Downsampling improved the performance of the classifier across all the models, the training cross validation results before and after the final iteration of downsampling can be seen in \Cref{tab:TrainingResults}. All of the results presented in this work are based on experiments conducted using the downsampled dataset. We have stated explicitly in case the entire dataset is used.

\subsubsection{Preprocessing and Vectorization of Error Messages}

\begin{figure}[ht]
 \caption{Example of Error Message preprocessing }
 \includegraphics[width=1.0\columnwidth]{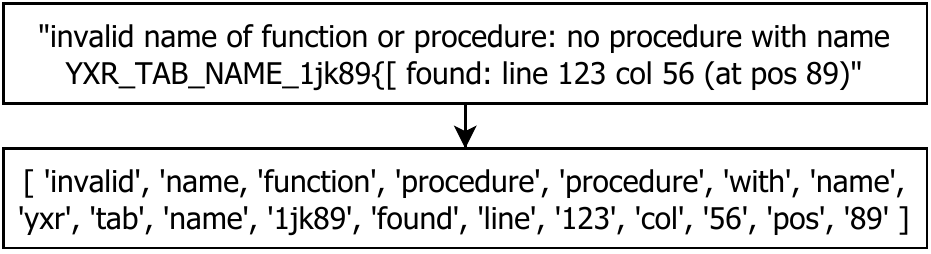}
  \label{preprocessedErrmsg}
\end{figure}

\begin{figure*}[h]
\caption{Data overview}
\vspace{3mm}
\centering
	\begin{minipage}[c][0.6\width]{0.4\textwidth}
	 \centering
    	 \includegraphics[width=0.73\textwidth]{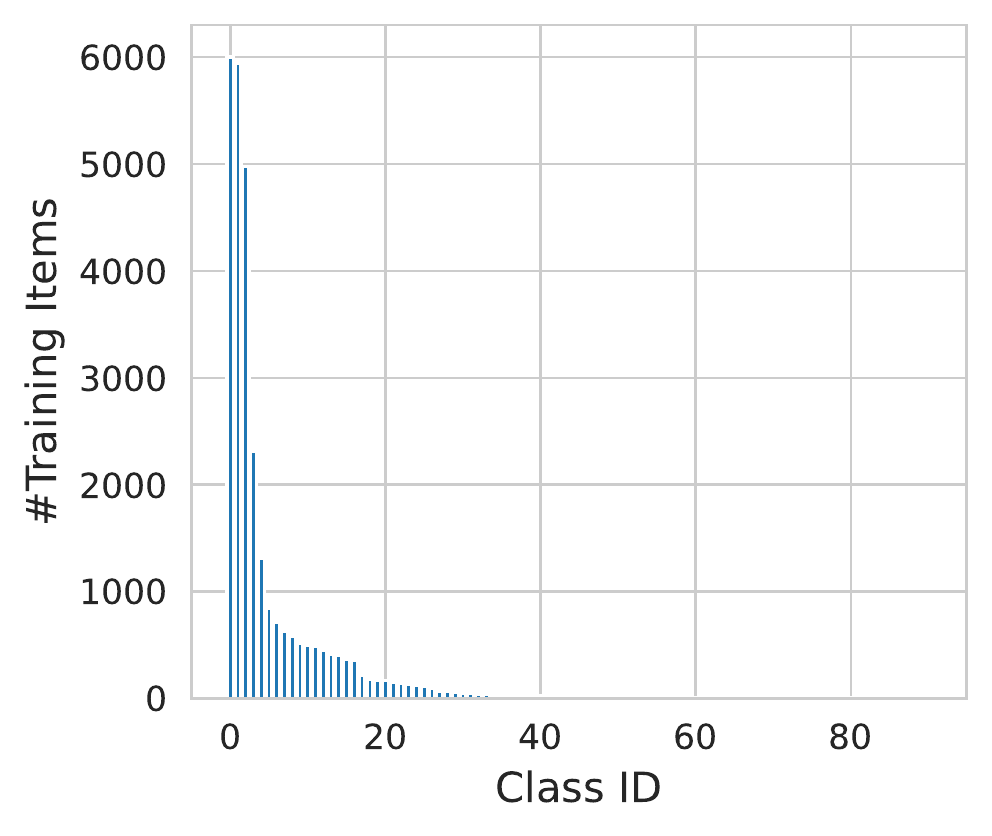}
	    \subcaption{Class distribution}
	 \label{fig:ClassDist}
	\end{minipage}%
	\begin{minipage}[c][0.6\width]{0.4\textwidth}
	 \centering
	 \includegraphics[width=0.73\textwidth]{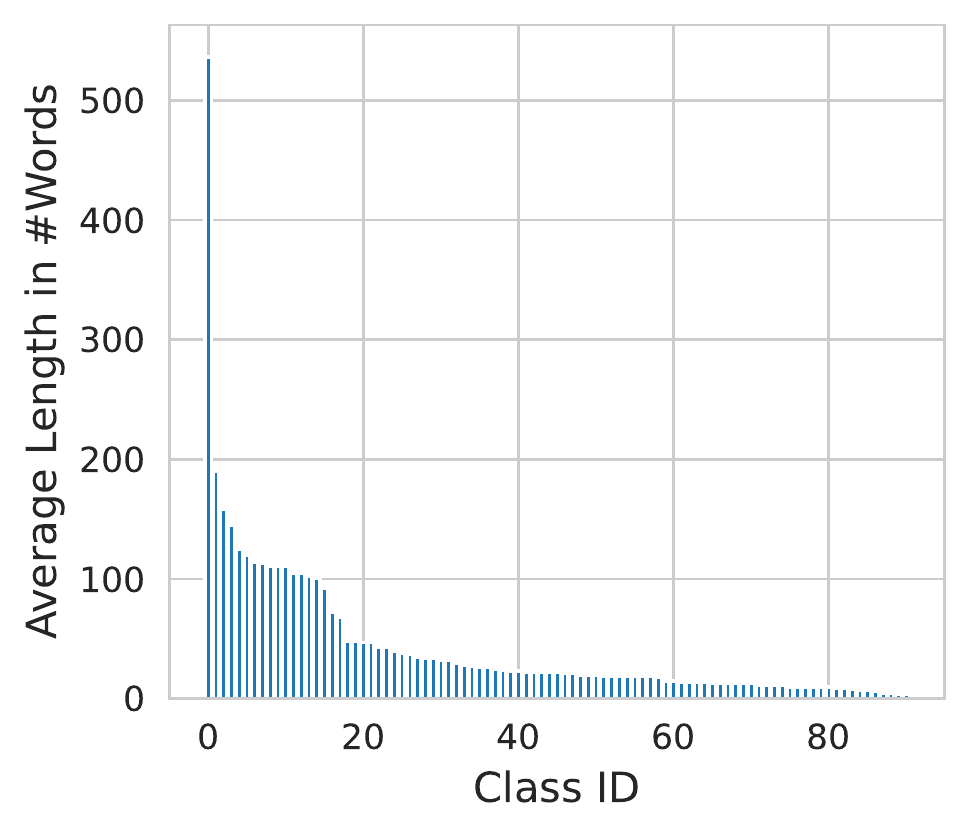}
	 	 \subcaption{Average length of error messages in each class}
   \label{AverageErrMsgLen}
	\end{minipage}
\label{fig:DataOverview}

\end{figure*}

 We present the average lengths of the error messages in each class in \Cref{AverageErrMsgLen}. Some of the error messages are very long but 90\% of our classes have error messages containing less than 100 terms. We utilize the domain knowledge of our experts to ascertain the contributions of error message patterns towards identification of the root causes. Error messages generally contain valuable information like location of the error which includes line numbers and positions. There can be errors coming from the same file but with different root causes. Retaining numbers present in the error messages helps us to exploit this information. However, not all numbers are equally useful, for instance, the error messages also contain temporary table names with numbers, alphabets and nonalphanumeric characters. Such temporary names are unique in each error message and are not useful for identifying root causes. Therefore, we attempt to discard such strings as a part of preprocessing the error messages.  Our final preprocessing steps involve lower casing, removal of nonalphanumeric characters, removal of stopwords.  

\Cref{preprocessedErrmsg} shows an example of a preprocessed error message. Five different classes share this error message pattern. Among these classes, differences are only with respect to the name of the table the error occurs on and the location of the error. The numbers appearing after terms like 'line', 'pos' also indicate a root cause in our data. Retaining the context of terms through word order is beneficial in such cases. A Bag of Words\cite{zhang2010understanding} model like TFIDF ignores the word order and therefore will fail to capture such dependencies. Since our Error Messages are short, creating $n$-Grams can help us to retain the word order but this will result in high dimensional data. Due to these reasons, we utilize the Doc2Vec to create embeddings for the error messages. We additionally tag each error message with the class label while training the embedding model. This helps us to establish an additional context of patterns belonging to the same root cause.  Terms such as '1jk89' may repeat only in very rare cases and are mostly eliminated by setting a suitable minimum term frequency during vectorization. Based on the hyperparameter optimization, 3 is a suitable minimum term frequency for both TFIDF and Doc2Vec models. We are able to retain all of the important terms from the minority classes and discard most of the unimportant terms from the vocabulary of the models. Doc2Vec  results in better performance across most of our classification experiments. Therefore, we productively use Doc2Vec as a vectorizer in MIRA. 

\begin{table}[ht]
\caption{Final hyperparameters from 5-fold cross-validation}
\label{tab:FinalHyperparameters}
\begin{tabular}{@{}lll@{}}
\toprule
Model     & Library & Final hyperparameters        \\ \midrule
TFIDF     & Sklearn\cite{scikit-learn} & stopwords='english', max\_df=1.0, min\_df=3   \\\cmidrule(l){3-3}
Doc2Vec &
 Gensim\cite{rehurek_lrec} &
 \begin{tabular}[c]{@{}l@{}}epochs=50, vector size=50, min\_count=3, \\ window=5, epochs for infer\_vector=200\end{tabular} \\ \cmidrule(l){3-3}
KNN+CD & Sklearn & K=11, weighted by distance, method='brute' \\ \cmidrule(l){3-3}
KNN+ED & Sklearn & K=11, weighted by distance, method='auto' \\ \cmidrule(l){3-3}
XGBoost &
 XGBoost\cite{Chen:2016:XST:2939672.2939785} &
 \begin{tabular}[c]{@{}l@{}}max\_depth=20, learning\_rate=0.1\\ objective='multi:softprob', booster='gbtree', \\ eval\_metric='logloss', n\_estimators=300\end{tabular} \\ \bottomrule
\end{tabular}
\end{table}

\subsubsection{Feature Selection}
 We conducted experiments to compare the performance of classification models trained only using vectorized error messages and models trained with all the attributes. We do not present results for KNN CD here as without additional attributes it is merely a KNN with cosine distance. Even though the F1-score from 5-fold cross-validation is presented in \Cref{tab:ResultsUsingErrMsgOnly} indicate a reasonable performance using only Error Messages, models built with all the selected attributes have better F1-scores across all the classifiers as presented in \Cref{tab:TrainingResults}. With these experiments our final selection of attributes includes Error Codes, SQL types, SQL sub types, Request Names, and Error Messages of which Error Message is the most important attribute.

\begin{table}[ht]
\centering
\caption{ Average F1-Scores 5-fold cross-validation on training data using only Error Messages }
\label{tab:ResultsUsingErrMsgOnly}
\begin{tabular}{@{}lcccc@{}}
\toprule
                     & \multicolumn{4}{c}{Classifiers}                                                                                      \\ \cmidrule(l){2-5} 
\multicolumn{1}{c}{} & \multicolumn{2}{c|}{F1 Weighted}                          & \multicolumn{2}{c}{F1}                                   \\ \cmidrule(l){2-5} 
Vectorization        & \multicolumn{1}{l}{XGBoost} & \multicolumn{1}{l|}{KNN+ED} & \multicolumn{1}{l}{XGBoost} & \multicolumn{1}{l}{KNN+ED} \\ \midrule
TFIDF                & 94.78                       & \multicolumn{1}{c|}{93.66}  & 91.03                       & 93.09                      \\
Doc2Vec              & 94.52                       & \multicolumn{1}{c|}{95.12}  & 92.40                       & 93.40                      \\ \bottomrule
\end{tabular}
\end{table}




\begin{table*}[ht]
\centering
\caption{Average F1-Scores of 5-fold cross-validation on training data}
\label{tab:TrainingResults}
\begin{tabular}{@{}llcccccc@{}}
\toprule
                             &               & \multicolumn{6}{c}{Classifiers}                                            \\ \cmidrule(l){3-8} 
                             &               & \multicolumn{3}{c|}{F1 Weighted}                & \multicolumn{3}{c}{F1}    \\ \cmidrule(l){3-8} 
Data                         & Vectorization & KNN+CD & KNN+ED & \multicolumn{1}{c|}{XGBoost} & KNN+CD & KNN+ED & XGBoost \\ \midrule
Downsampled & TFIDF         & 99.13  & 95.67  & \multicolumn{1}{c|}{98.78}   & 98.69  & 98.20  & 98.31   \\
                             & Doc2Vec       & 99.70  & 98.98  & \multicolumn{1}{c|}{99.41}   & 98.80  & 97.63  & 97.96   \\ \midrule
Full data                     & TFIDF         & 98.12  & 94.32  & \multicolumn{1}{c|}{96.67}   & 97.83  & 95.01  & 97.01   \\
                             & Doc2Vec       & 98.67  & 96.50  &  \multicolumn{1}{c|}{97.70}                        & 97.01  & 95.32  & 97.21   \\ \bottomrule
\end{tabular}
\end{table*}

\subsection{Analysis of the Classification Results}
 We analyse the performance of the machine learning component in classifying failed replay events. We discuss our findings from the analysis of cross-validation results of training data and also the model performance in our production environment. \Cref{EvalOverview} presents an overview of the process.

\subsubsection{Cross-Validation Results of Training Data}
Despite our experts' label verification efforts, our full training dataset with \num{65000} samples had some incorrectly labeled events. Since some of the misclassifications revealed the wrongly labeled events, we identified and corrected the remaining labeling discrepancies during our initial cross-validation with training data. This process also compelled a more thorough analysis of the root cause categorization in our data and helped us restructure the root causes into more meaningful clusters than before. Most of the remaining misclassifications are due to events belonging to classes that share all values except the error message. The error messages from these classes also share a lot of vocabulary. E.g., similar errors with different table names: \textit{'Cannot find the table name <TableName>'}. We could resolve many of these issues by continuously adapting the text preprocessing for error messages and selecting an appropriate vectorization technique, in our case Doc2Vec.

 As seen in our experiments, KNN with CD has a better F1-score in most cases compared to our baselines. XGBoost performs on par with our approach and is also a comparatively faster classifier than KNN for predictions. However, for employing the classification model in production, we analyze the results with a focus on two aspects. i) the model should be able to reflect uncertainties when the failed events cannot be classified with high certainty. ii) interpretability of the results. We set the thresholds for the probability based on the analysis of correctly classified events in the cross-validation results. There are on average only 11 events per fold that are correctly classified but have probabilities below the threshold of 0.9 in case of KNN with CD. In case of XGBoost, there are on average 110 correctly classified events whose probabilities ranged from 0.5 to 0.8 and are below the threshold setting of 0.8. The XGBoost model will result in a relatively larger workload for operators in comparison to the KNN with custom distance model. Furthermore, there are more possibilities of uncertain classifications surpassing a lower threshold setting for a certainty measure, which we attempt to avoid in MIRA. It is currently harder to set such a threshold in case of XGBoost. The prediction probabilities of KNN with CD have less variance in comparison to the probabilities of XGBoost. Furthermore, the classifications of KNN models are easier to interpret than the classifications of XGBoost. This is especially true when Doc2Vec is used for vectorization. MIRA logs the neighbors and their distances during a classification, making it easy to analyze the results. For these reasons, we are currently using the KNN with CD to classify failed events in MIRA. However, in our case, XGBoost is a candidate for creating an ensemble setup in MIRA.

 \subsubsection{Retraining} We currently monitor the classification results very closely in MIRA. The operators comprehensively analyze all uncertain classifications to identify misclassifications and correct them and add them to a collection. In this collection  we also include the new failure types identified through an ongoing manual replay analyses. We augment the training data with this collection of samples and retrain the classifier once a week. Based on our observations, retraining has improved the performance in the subsequent classification results in terms of accuracy and confidence of the predictions. 

\subsubsection{Analysis with Visualizations}
\begin{figure}[ht]
\hfill
 \caption{Visualizations of Doc2Vec embeddings using t-SNE}
 \includegraphics[scale=1.0]{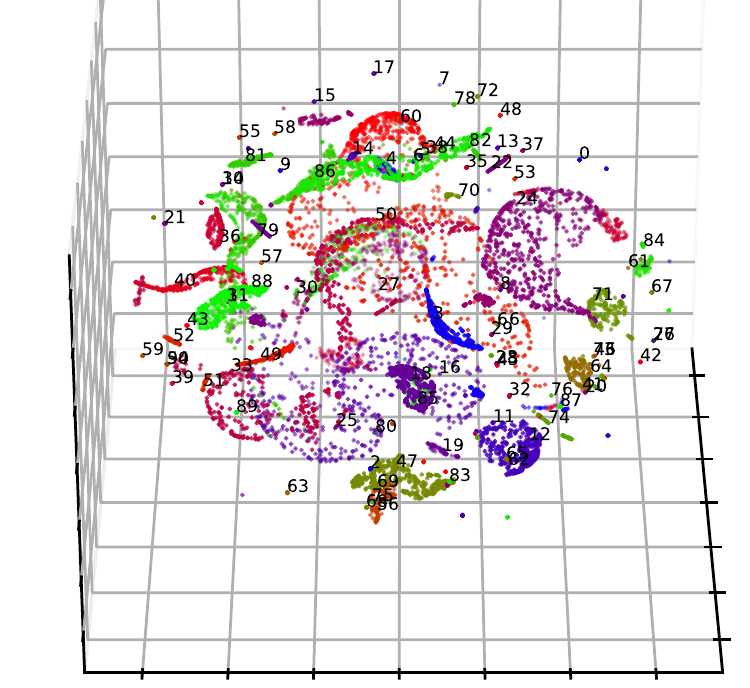}
  \label{TSNEEmbeddings}
\end{figure}

 Considering that the Error Messages play the most important role in our classification process, obtaining an approximation of how the error messages of different classes are embedded in the vector space with respect to each other is helpful. We visualize the Doc2Vec embeddings using the t-SNE \cite{van2008visualizing}. In \Cref{TSNEEmbeddings} we present a brief overview of the embeddings of a subset of our data. The label and colors of the clusters indicate different classes. Some of the classes, like 58 and 36, have dense and well separated clusters from other classes. Such classes have less variance in their error patterns and are generally not misclassified. Clusters that are larger and more dispersed within a single class suggest the presence of multiple error patterns. Classes such as 27 and 50 share vocabularies and are prone to misclassifications.

 In practice, the visual guide is helpful to the operators as well to obtain an overview of the proportion of the classes in the data, identify the categories of root causes that are placed very close to each other and prone to misclassification. Currently the MIRA administrators make a final decisions on many aspects such as adding new training items and accepting corrected classifications. As the operators gain more experience with MIRA, they can rely on such visualizations to make the aforementioned decisions, making this process more democratic. Based on our discussions with the operators, an interactive visualization for the Error Message embeddings along with a metric like F1-Score to indicate the changes in performance from one training iteration to the next, can be beneficial to understand the classification results of MIRA.

\subsubsection{Alternative approaches}

We evaluated an alternative approach by concatenating all the attribute values for an event into one string and vectorizing it to use as a input to the classifiers. The average F1 Score over 5 fold Cross-Validation is shown in \Cref{tab:ConcatenatedResults}. This approach impacts the classification of several minority classes. However for our data, this approach is a promising alternative.


\begin{table}[ht]
\centering
\caption{ Average F1-Scores 5-fold cross-validation on concatenated and vectorized features}
\label{tab:ConcatenatedResults}
\begin{tabular}{@{}lcccc@{}}
\toprule
\multicolumn{1}{c}{} & \multicolumn{4}{c}{Classifiers}                                                                                      \\ \cmidrule(l){2-5} 
\multicolumn{1}{c}{} & \multicolumn{2}{c|}{F1 Weighted}                          & \multicolumn{2}{c}{F1}                                   \\ \cmidrule(l){2-5} 
Vectorization        & \multicolumn{1}{l}{XGBoost} & \multicolumn{1}{l|}{KNN+ED} & \multicolumn{1}{l}{XGBoost} & \multicolumn{1}{l}{KNN+ED} \\ \midrule
TFIDF                & 99.11                       & \multicolumn{1}{c|}{98.32}  & 95.13                       & 95.32                      \\
Doc2Vec              & 98.78                       & \multicolumn{1}{c|}{99.20}  & 94.61                       & 93.12                      \\ \bottomrule
\end{tabular}
\end{table}

\subsubsection{Performance of MIRA in Production Environment}
We discuss the performance of MIRA in production in two phases i) \textit{Phase 1}: Performance of model trained on the full dataset  ii) \textit{Phase 2}: Performance of the optimized model based on the feedback from Phase 1. \Cref{EvalOverview} presents an overview of the two phases. Using MIRA, we have classified failed events from 51 replays from 7 different captured workloads. The number of failed events per replay ranged from \num{3000} to \num{161000}, with 4 to 35 classes per replay. We found 75 unique classes out of 93 classes present in the training data. Common settings across both the phases include Doc2Vec model for the Error Messages, a trained KNN with CD, threshold of 0.9 for probability and 0.7 for confidence to filter uncertain classifications. We closely examined MIRA's classifications of failed events from at least 28 replays where the reasons for failures were known. 
Whenever the predicted reason is correct but one or both of the certainty measures fall below the defined thresholds, we consider this a false uncertainty. Due to the fact that operators investigate all uncertain classifications, false uncertainties are regarded as additional workload caused by MIRA. As a result, we use false uncertainty as one of the metrics to evaluate MIRA. \Cref{tab:ProdEvalResults} summarizes our production evaluation.

In Phase 1, we utilized the model trained with the full dataset with about \num{65000} training samples with K=3 for KNN. We set a low K value to accommodate the minority classes. We classified failed events from 24 replays, of which we describe the analysis of 2 replays belonging to the same capture. Each replay has over 13,000 failed events and four root cause classes. Classification accuracy for Replay 1 was \SI[round-mode=places,round-precision=2]{97.83}{\percent}. In Replay 1, uncertainty measurements correctly identified the misclassified events which are then reclassified and added to the training data to retrain the classifier. With the retrained classifier, classification accuracy for Replay 2 was \SI[round-mode=places,round-precision=2]{99.4}{\percent}. Since both the replays belong to the same capture, retraining the model with corrected classifications improved the classification accuracy for Replay 2. There were less than 0.5\% false uncertainties across both replays. However, subsequent replay classifications deteriorated in terms of accuracy and certainty, resulting in over 15\% false uncertainties, increasing the analysis workload. We discovered that the majority classes hampered the predictions of several minority classes, and the low K value exacerbated the problem. As a result, we downsampled the majority classes, as indicated in \Cref{classimbalance}, and further optimized the model.

In Phase 2, the model was trained on downsampled data with \num{25600} samples, restructured root causes and K=11.  \Cref{fig:DataOverview} shows an overview of the final dataset. The classifications are comparatively more robust, with a higher probability and confidence for correct classifications, based on a detailed analysis of 10 replays. A smaller number of training events results in fewer distance calculations and a faster model. The average prediction accuracy is about 97\%, with less than 0.5\% false uncertainties across all 10 replays. Most of the uncertain classifications in MIRA are misclassified events that pertain to an untrained new root cause category.


 Across both the phases, despite the need for analysis of results, frequent changes and updates to the model, we have verified that MIRA results in a significantly smaller workload for the operators in comparison to the previous approach of replay analysis for SAP HANA. 

\begin{table}
\centering
\caption{Average accuracy and false uncertainty from evaluation of MIRA in production}
\label{tab:ProdEvalResults}
\begin{tabular}{@{}llllll@{}}
\toprule
Phase & \#Replays & Accuracy & False Uncertainty &  Data        & KNN  \\ \midrule
1     & 18       & 98.61         & 0.5\% to 15\%       & \num{65000}   & K=3  \\
2     & 10      & 97.83         & \textless{}0.5\%    & \num{25600} & K=11 \\ \bottomrule
\end{tabular}
\end{table}

\section{Practical Application and Discussion}
\label{s:ApplicationDiscussion}
In this section, we describe the contributions of MIRA towards replay analysis at SAP, expand on some of the challenges we encountered in the process of applying MIRA in practice and propose possible solutions. We briefly discuss the feedback from the end-users of MIRA.

\subsection{Application of MIRA}
MIRA does not depend on a particular capture and therefore allows a high degree of automation for result analysis in comparison to the existing rule-based approach. It improves the user experience for the operators because it reduces manual work and the time between test result and test assessment. Due to categorization of the failed events into known issues, the operators have a much smaller workload of only analyzing uncertain results from MIRA. MIRA serves as a central repository for all known issues and eliminates having to maintain error metadata per capture. Ultimately, MIRA offers a much more scalable analysis and allows us to scale the entire capture and replay process to many concurrent systems as part of the HANA test and release pipeline.

Employing MIRA gives us a holistic overview of all product issues in replays, their frequency, and distribution across customer systems. This permits tracking of bugs over time and code line. This leads to better product release and upgrade decisions, as well as customer-specific optimizations like configuration and data distribution changes.
 


\subsection{Feedback from End-Users}
The productive usage of MIRA provides an immediate benefit to operators, as database issue are pre-analyzed by the machine learning model to classify the symptoms to the correct root cause. This greatly helps to synchronize analysis efforts as symptoms are not analyzed multiple times if they appeared on multiple replays. With our experiments, MIRA has shown reliable performance across several replays. Therefore the operators are not required to remember a variety of different symptoms and their actual root cause as they can trust the initial assessments and classifications of the model.

However, the introduction of a machine learning tool, which relies on a constant feedback and training loop from operators, has shown that this also creates an overhead for operators, as they have to analyze the machine learning predictions and correct them if needed. These additional efforts have to be considered, especially in phases where the model is still in its early stages of training and configuration. Ultimately, as data from more replays is gathered and trained into the model, we assume that this operational overhead will gradually decrease over time. To ensure that the amount of manual checks and interventions is limited to a minimum, multiple conditions have to be provided by MIRA: i) Ensure a stable classification result over multiple predictions to avoid operator confusion and distrust in the model's judgement. ii) Provide a confidence metric that reliably indicates uncertainties to an operator to trigger a manual intervention. iii) Ensure that the model is re-trained continuously to represent the new HANA errors as quickly as possible in the subsequent predictions.

\section{Challenges, Limitations and Improvements}
\label{s:challenges}
 We identify new failure types frequently in an ongoing replay analysis and therefore the data for MIRA also evolves. As the data for learning in MIRA grows and diversifies, we will encounter problems that will need our current setup and the models to be constantly adapted. In this section we briefly discuss some of the challenges we faced in MIRA and report limitations and potential improvements of our solution.

\subsection{Challenges of Evolving Data}
The evolving data leads to two main challenges: i) Identical Failed Events, and ii) Change in Structure of Table Names.

\textit{i) Identical Failed Events:}
Historically, system faults in HANA have been propagated transparently to administrators and customers. Although this gives a lot of context for failures for expert operators, it also creates ambiguity for end users as the errors are hard to understand. Recently, components are partially unifying and standardizing the way they convey system failures to end users. In practice, components now offer less error context and more guidance for customers or end users on how to deal with an issue. This category of failures has fixed error messages. For instance, \textit{"further analysis is required to identify the root cause of the failure"}. Since these failures produce identical events, MIRA cannot classify them using the current set of features. Uncertainty measures will fail to identify them as well.
\begin{figure}
 \includegraphics[scale=0.83]{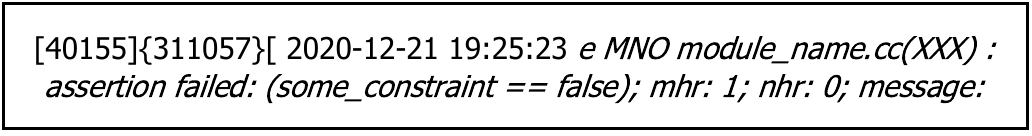}
  \caption{Example of an artificial trace file entry of Failed Assertion: Error messages are augmented with italicized parts}
  \label{StackTrace}
\end{figure}

In some of these cases, additional error context exists that helps the operators classify the errors correctly. Stack traces are one of the event attributes available for specific HANA failures. In these cases, the operators manually analyze the stack traces collected for the failures to distinguish between such failures. With an initial analysis, we were able to identify helpful patterns within the trace files across all such failures to help the model distinguish between such events. For instance, we present a sample entry of an assertion failure in \Cref{StackTrace}. We preprocess the entries in the trace files to extract an entry carrying such an error message and append this extracted string to the original error message. These augmented error messages are then used as input to create the Doc2Vec embeddings for the error messages. Our data contains several categories of failure, and trace files are not available for all of the failure classes. Whenever a trace file is available for the failure, we extend the preprocessing steps for the error messages.

\begin{figure}[ht]
\hfill
	\begin{minipage}[c][0.66\width]{0.485\textwidth}
	 \subcaption{Without using trace files}
	 \centering
	 \includegraphics[scale=0.62]{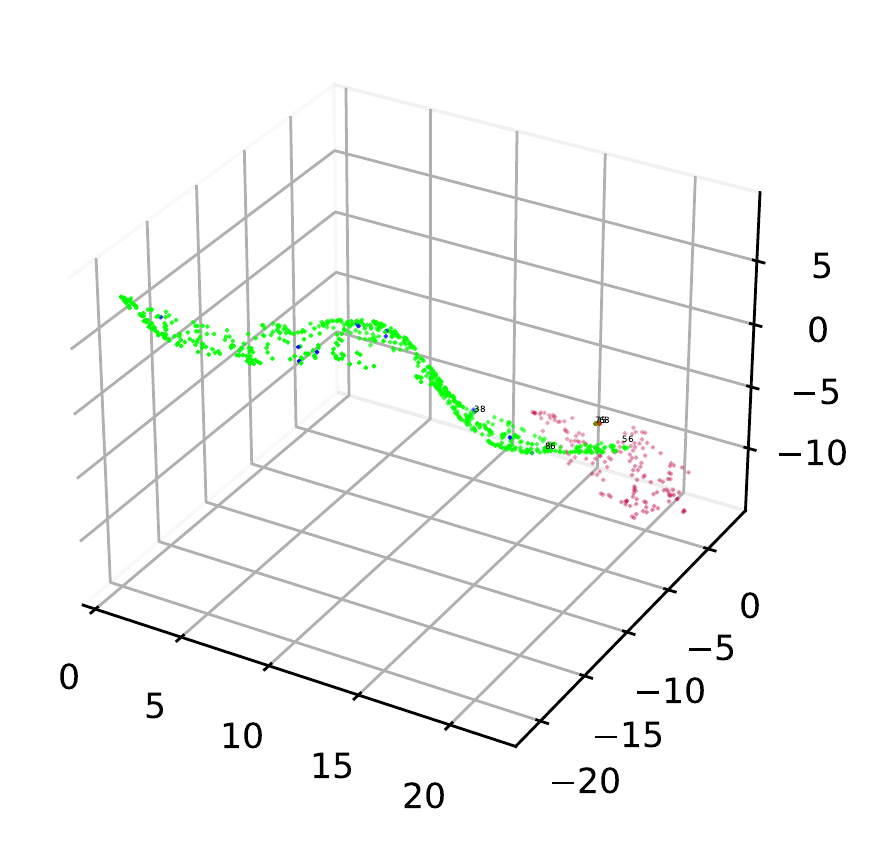}
	 \label{NoTrace}
	\end{minipage}
	\begin{minipage}[c][0.64\width]{0.485\textwidth}
	 \centering
	 \includegraphics[scale=0.62]{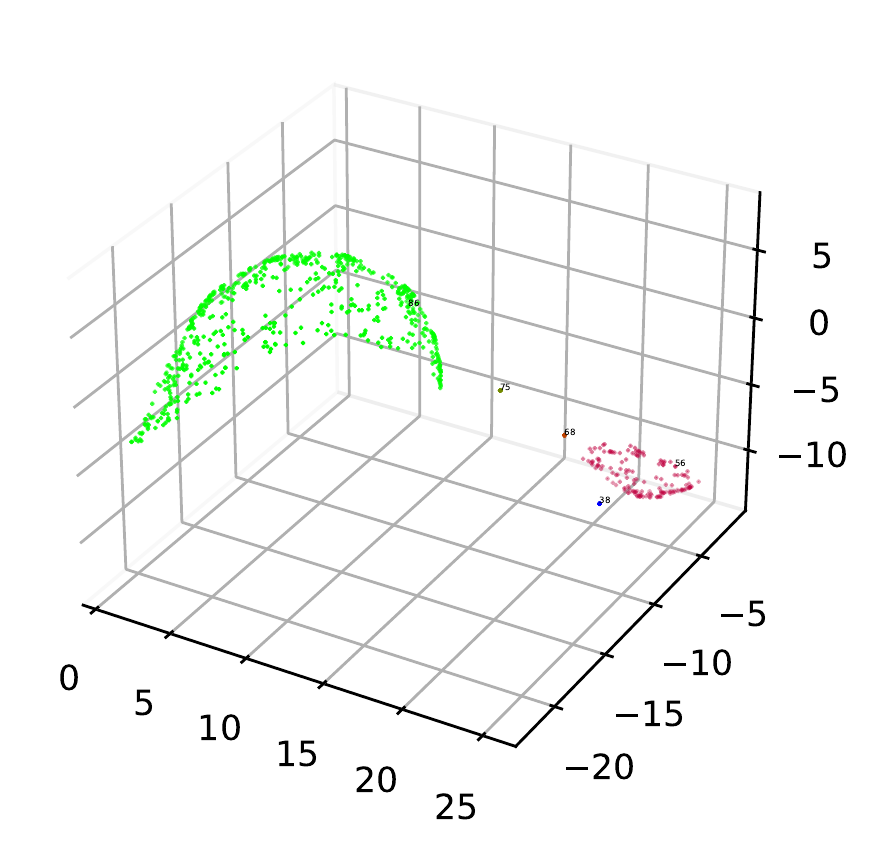}
	 	 \subcaption{Using trace files}
   \label{WithTrace}
	\end{minipage}
	\caption{Comparison of changes in Doc2Vec embeddings before and after using trace files}
	\label{TraceEmbeddings}

\end{figure}

 We performed a 5-fold cross-validation on the training data to test this approach. We were able to successfully classify this category of failed events into their correct root causes. We present the change in Doc2Vec embeddings using t-SNE plot in \Cref{TraceEmbeddings} where the classes appear to be well separated after using the trace file entries. We currently only have 5 classes with such a scenario. The frequency of such failures is rare. We need to gather more data and conduct more experiments to apply this solution productively. However, the stack traces prove to be a valuable feature for classifying replay failures.

\textit{ii) Change in Structure of Table Names:} 
The table name is the most important part of the error message for identifying several root causes in MIRA. In our data, the table names with dynamically generated parts mostly have a structure of 'TABLENAME\_\textit{\textbf{ijh78fk}}' where the dynamic part is concatenated to the table names with an '\_'. This is easily separated and eliminated during the text preprocessing steps for error messages. However, in some of the recent errors, we have discovered instances with table names like 'TABLENAME\textit{\textbf{ijh78fk}}'. This presents new challenges to our existing set up, like the minimum frequency setting for text vectorization and the vectorization technique itself. Due to the unique temporary string, these terms will always be treated as out-of-Vocabulary (OOV) terms and therefore will have no influence on the classification. The model will be severely handicapped by such OOV issues. To alleviate these issues, we have identified fasttext\cite{bojanowski2017enriching} as an alternative for vectorizing the error messages. fasttext creates embeddings for the subword vectors and therefore can generate vectors for such OOV words by using a subword like 'TABLENAME'. 

\subsection{Limitations and Improvements}
The data will change over time when new software defects are introduced. It is a challenge to recognize changes and to maintain an up to date training data, retrain the classifier and reassess the threshold settings for the model certainty. Both the text vectorization techniques, TFIDF and Doc2Vec, require retraining to accommodate the vocabulary of new failures. Additionally, document embeddings may need retraining when there is a remarkable change in context of terms in the new data in comparison to the data used for learning the embeddings.

The aspect of detecting changes can also be generalized to the general question of how machine learning approaches are tested from the software engineering perspective. As machine learning applies statistics, it inherently has some uncertainty about the results. It remains unclear, and to our knowledge of the current state of research, it is also an unsolved problem, how the correctness of machine learning approaches should be tested. This results in several practical questions such as: What is the meaning of a test oracle in the machine learning domain? What exactly is a failure for a machine learning result? What metrics can we use for test adequacy? We welcome any ideas and suggestion towards these questions.
 
Currently, MIRA retrains the classification model with corrected classifications, creating a semi-supervised learning setup. This setup is still prone to issues of inconsistent correction of uncertain classifications by operators and misclassifications not captured by both the uncertainty measures. Such events getting trained into the model will lead to deterioration of the model quality over time. MIRA administrators are responsible for observing the model quality and taking necessary action. 

\section{Conclusion}
\label{s:conclusion}

In this work, we discussed MIRA, an approach to automatically classify failed replay events into their root causes using machine learning. As a part of the solution, we propose a customized distance function to mitigate issues of encoding categorical attributes containing several values resulting in high dimensional data. We have shown our approach to be highly accurate and reliable with our evaluation for our data. MIRA makes the process of replay failure analysis more consistent, saves manual effort and allows to focus on solving issues concerning the quality of the underlying software. The involved effort incurs considerably lesser cost than the error-prone manual inspection. We also highlighted several practical limitations and open questions in the areas of supporting changes over time, testing machine learning based approaches, and best practices for human-in-the loop architectures.

We intend to automate model monitoring and retraining within MIRA in the future. Considering that we find stack traces useful for categorization, we're working on adding an exception stack trace comparison as an additional feature and expanding MIRA's custom distance computation.

\bibliographystyle{IEEEtran}
\bibliography{references}

\end{document}